\documentclass[prfluids, amsmath, amssymb, aps]{revtex4-2}

\usepackage{graphicx}

\usepackage{siunitx}

\usepackage[T1]{fontenc}

\usepackage{xcolor}
\begin{document}

\title{Multi-scale energy budget of inertially driven turbulence in normal and superfluid helium}

\author{Fatimata Sy}
\author{Pantxo Diribarne}
\author{Bernard Rousset}
\email{bernard.rousset@cea.fr}
\affiliation{dSBT/IRIG CEA, Universit\'{e} Grenoble Alpes - F-38054 Grenoble, France}
\author{Mathieu Gibert}
\affiliation{Institut NEEL, CNRS, Universit\'{e} Grenoble Alpes - F-38042 Grenoble, France}
\author{Mickael Bourgoin}
\affiliation{LEGI, CNRS, Universit\'{e} Grenoble Alpes}

\begin{abstract}
In this paper we present a novel hydrodynamic experiment using liquid $^4$He. The flow is forced inertially by a canonical oscillating grid using either its normal (He~I) or superfluid (He~II) phase, generating a statistically stationary turbulence. We characterise the turbulent properties of the flow using 2D Lagrangian Particle tracking on hollow glass micro-spheres. As expected for tracer particles, the Vorono\"{i} tessellation on particle positions does not show a significant departure from a random Poisson process neither in He~I nor He~II phase. Particles’ positions are tracked with high temporal resolution, allowing to resolve velocity fluctuations at integral and inertial scales while properly assessing the noise contribution. Additionally, we differentiate the particles’ positions (by convolution with Gaussian kernels) in order to access small scale quantities like acceleration. Using these measured quantities and the formalism of classical Homogeneous Isotropic Turbulence (HIT) to perform an energy budget across scales we extract the energy injection rate at the large scale, the energy flux cascading through inertial scales, down to small scales at which it is dissipated. We found that in such inertially driven turbulence, regardless of the normal or superfluid state of the fluid, estimates of energy at the different scales are compatible with each other and consistent with oscillating grid turbulence results reported for normal fluids in the literature. The largest discrepancy shows up at small scales where the signal to noise ratio is harder to control and where the 2D measurement is contaminated by the 3D nature of the flow. This motivates to focus future experimental projects towards small scales, low noise and 3D measurements.

\end{abstract}

\maketitle

\section{\label{sec:intro}Introduction}
Liquid Helium experiments offer a unique way to investigate developed turbulence in laboratory scale facilities.
Liquid Helium, in its normal state (He I), has indeed a very low
kinematic viscosity $\nu$. 
Besides, one of the most striking features of liquid Helium is
superfluidity (He~II) where the kinematic viscosity eventually vanishes below
a critical temperature~$T_{\lambda}~\approx~2.17~K$.

While He~I follows a classical Navier-Stokes equation dynamics for a
viscous newtonian fluid, He~II is usually described as a mixture of a
normal (viscous) and a superfluid (inviscid) components with a relative
fraction depending on the temperature (the lower the temperature the
higher the superfluid fraction). A consequence of the inviscid nature
of the superfluid component is that turbulent eddies in that component
cannot have arbitrary  circulation: only vortices carrying a single
quantum of circulation  $\kappa$ may exist, the  so-called quantum
vortices~\cite{Vinen2002}. Those vortices act as defects where the
excitations from the normal component may scatter. This mechanism
produces a transfer of momentum between the two components of He~II
leading to a mechanical coupling called mutual friction.

At finite temperature and  in absence of temperature gradient, it is
believed that this mutual  friction locks the velocity fields of the
two components at scales  larger than the typical inter-vortex
distance  $\delta$. 
That explains the lack of observed difference in energy spectra
at large scale~\cite{Maurer98, Salort10}. At scales comparable to
$\delta$, it is predicted that the mutual friction is not strong
enough to lock the two components together hence a different behaviour
compared to  classical turbulence is predicted~\cite{Skrbek06}.
The inter-vortex distance is expected to be of the same order of
magnitude as the dissipative scale in the normal component, which may
be very small in laboratory experiments (typically ranging between one and
several tens of micrometers). \\\\

Since Eulerian sensors are difficult to use in He~II and do not even
exist at scales of less than a tenth of a millimetre, the use of
visualization to probe the flow has been explored in the past
decades~\cite{Chopra57, Kitchens65} as a promising approach
to access quantitative  multi-scale diagnosis of quantum turbulence.

$H_2$-$D_2$ ice particles and hollow grass spheres have initially been
used to assess the flow field in counter-flow
experiments~\cite{Murakami89, Bewley08, Paoletti08}.
Those reveal that Lagrangian statistics at small scales appear to
behave differently from those of the conventional fluid
\cite{LaMantia12, LaMantia16}. Nevertheless, 
due to the nature of the counter-flow itself (which has no counterpart
in classical fluid), and also to the small level of turbulence
involved in this situation, no clear conclusion has emerged yet
regarding possible intrinsic differences in the dynamics of super- and
normal-fluid turbulence when driven in similar conditions.

More recently, particle tracking velocimetry (PTV) has also been used
to measure statistical properties of inertially driven flows.
\textcite{Svancara17} used $H_2$-$D_2$ ice particles to look at velocity
and acceleration probability density functions in an oscillating grid
experiment. They found that the velocity and acceleration
distributions were comparable to that observed in standard fluids, as
already observed in Eulerian framework for scales larger than $\delta$.
\textcite{Tang20} studied the velocity  structure
functions scalings in a towed-grid experiment, in He~II only.
They conclude, among other, that they observe a larger intermittency
than in classical fluids, on the basis of comparison with theoretical models. 

It remains unclear at the moment which component the Lagragian
particles actually trace in HeII (see e.g. Ref.~\cite{Mastracci18}).
One goal of the present study is to proceed to different estimates of
energy across scales in order to explore possible deviations to
classical behaviors, which may indicate any specificity of superfluid
behavior (due either to a preferential 
sampling of the tracer to one component or the other, or to the
existence of different channels for energy to flow and dissipate
across scales). To this end, we have estimated the energy rates at
different scales, always assuming fundamental laws as they are known
for classical fluid turbulence, seeking scale by scale for significant
differences between measurements carried in HeI and HeII.  This
direct comparison is the most reliable way to highlight features
peculiar to He~II.

To achieve this, we designed a new experimental facility devoted to
particle tracking and particle trapping measurements in Oscillating
Grid Turbulence~\cite{SY15} (OGT). The main difference with
\textcite{Svancara17} is that we chose to follow the design rules of a
canonical oscillating grid experiments, e.g. grid solidity below 40\%
and at least 3x3 meshes with half mesh at each end\cite{Hopfinger1976,
  Fernando93, Fernando94}.  Thereby our experiment offers the
possibility of calibrating and validating our measurements in He~I
against classical reference data.
Compared to grid generated turbulence in wind tunnels, OGT has
the advantage to produce a flow with almost no mean flow, hence better
suited to particle tracking experiments with fixed cameras. Towed grid
experiments, which are now common tools to investigate inertially
driven turbulence \cite{Stalp99, White02, Tang20}, produce nearly
homogeneous isotropic flows but decaying (non-stationary) in time.
The choice of OGT for the present study was therefore motivated by its
stationarity. It is an important condition as it
ensures that energy injection rate, the energy transfer across
scales and the energy dissipation rate should compare equal.\\

In order to have a good control of particle size dispersion, we
decided to use hollow glass microspheres. Such particles have a
diameter of a few tens of 
micrometers though (an order of magnitude bigger than ice particles)
and part of our study aims at probing the relevance to consider such
particles as good tracers.\\

In section~\ref{sec:setup} we present the experimental
device. Section~\ref{sec:procedure} describes the typical experimental
protocol to optimise the operation of the facility for particle
tracking measurements. Section~\ref{sec:detection} describes particle
detection methodology before the exploration of particle trapping in
section~\ref{sec:trapping}. Section~\ref{sec:lpt} is dedicated to
particle tracking. Finally we present the velocity field in
section~\ref{sec:velfield} and section~\ref{sec:epsilon} is devoted to
the assessment of the dissipation rate at different scales of the
flow.  

\section{\label{sec:setup}Experimental setup}
In order to generate inertially driven turbulence in liquid Helium,
we designed an oscillating grid experiment together with a dedicated
cryostat. The final scope is to access Lagrangian velocity statistics
by means of particle tracking experiments in He~I and He~II and also
to investigate eventual preferential concentrations of particles in He~II.

\begin{figure}[ht!]
  \begin{center}
    \includegraphics[]{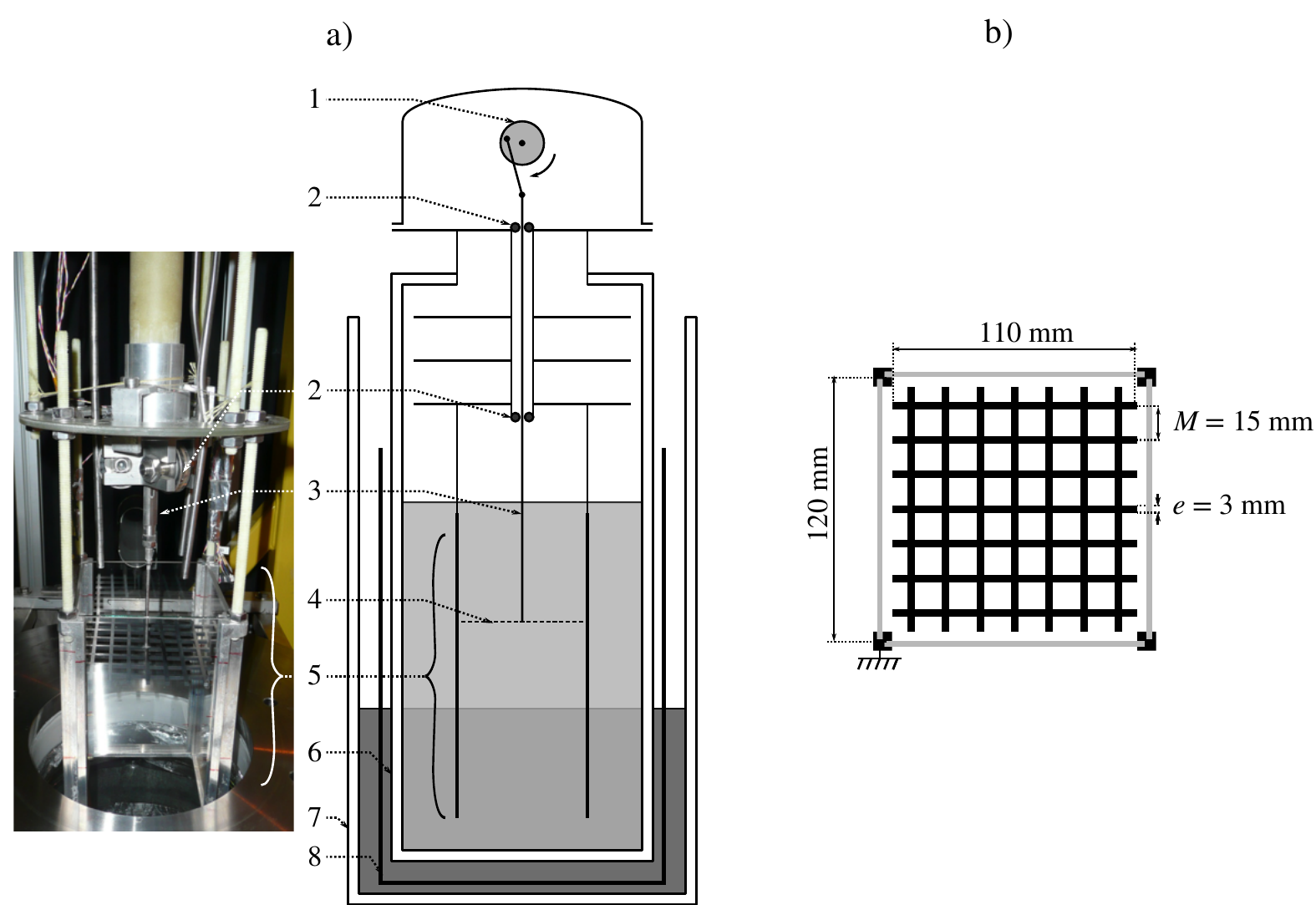}
  \end{center}
  \caption{a) Experimental setup: 1- crankshaft + gear-motor system, 2-
    ball bearing, 3- composite main shaft, 4- grid, 5- test section
    (``aquarium''), 6- inner (Helium) double wall glass vessel, 7-
    outer (Nitrogen) double wall glass vessel, 8- Aluminum radiation
    shield. 
    b) Details of the grid and test section dimensions}
  \label{fig:PresOgres}
\end{figure}

Fig.~\ref{fig:PresOgres}-a presents a simplified sketch view of the
experiment with the main elements of the facility.

In the sequel we will first detail the turbulence generation system,
then describe the cryostat and the visualization setup, and finally 
the particle seeding technique.

\subsection{Turbulence generation}
Turbulence is produced by oscillating a grid (item 4 in a
fig.~\ref{fig:PresOgres}-a) in a liquid helium bath (light gray). The grid is
driven by a gear motor (1), inside the cryostat but at room temperature,
via a shaft (3) which is designed to minimize heat losses due the thermal gradient. 

The motor is a MDP EC40 equipped with a planetary gear head GP 42 C
with a reduction ratio of 26:1. The system has a maximal rotation
frequency of \SI{6.4}{Hz} and can deliver a maximal torque of \SI{4.2}{N.m}. During
the experiments presented in this paper, the grid was always driven at 
constant frequency $f = \SI{5}{Hz}$.

We use a crankshaft with an adjustable stroke $S$ to convert the rotation to
quasi-sinusoidal vertical translation. The maximal stroke is about
30~mm, but experiments exposed in this paper were all performed with
$S=1.77\,M\approx \SI{26}{mm}$.

The grid oscillates vertically in a glass box (item 5 in
fig.~\ref{fig:PresOgres}-a) with square cross-section immersed in 
the bulk of liquid Helium.
The sides of the box are $W=\SI{120}{mm}$ large (marginally
larger than the grid itself, see fig.~\ref{fig:PresOgres}-b)
and the height is $H=\SI{250}{mm}$. The top and bottom ends of the box
are open: the goal of this "aquarium" is to ensure a
reproducible turbulence generation region, with well controlled
boundary conditions. The open top and bottom help in  minimizing
recirculating flows, although some residual large scale mean
recirculations are known to be hardly avoidable in oscillating grid
experiments. The four side walls of the aquarium are made of glass for optical
access purposes.

\subsubsection{Grid Geometry}

Fig.~\ref{fig:PresOgres}-b shows the grid geometry. It has been
designed based on  previous studies in classical fluids in order to
respect canonical  conditions on the solidity and the end
conditions~\cite{Fernando93}, known to produce a well characterized
turbulence, with good homogeneity and isotropy properties.

The grid is made of anodized aluminum with square bars and has a
solidity $G\approx36\%$. As a reminder, the solidity $G$ is defined as
the ratio between the frontal area effectively blocked by the bars and
the total cross section area of the grid, which for a square grid as
ours can be simply related to the bar width $e$ and the mesh size $M$: 

\begin{equation}
	G = \frac{e}{M}\left( 2 - \frac{e}{M} \right).
\label{eq.solidite}
\end{equation}

When the grid solidity exceeds 40\%, the jets and wakes produced by
the oscillation of the grid are known to become unstable and merge
together to form larger structures~\cite{Hopfinger1976}. We have
chosen a solidity of 36\% ($e=\SI{3}{mm}$, $M=\SI{15}{mm}$) for which "the wakes
coalesce with each other without bending their axes; shear-free
turbulence can be expected on either side of the grid at sufficiently
large z"~\cite{DeSilva1992}. 

Table~\ref{tab:caracgrid1} summarizes the grid characteristics.

\begin{table}[ht!]
    \centering
   \begin{tabular}{|c|c|c|c|c|} 
   \hline
   \multicolumn{5}{|c|}{\bf Grid}\\
   \hline
 $M$ & $e$ & $G$ & $f$ & $S$ \\
   \hline
[mm] & [mm] & [\%] & [Hz] &  [mm] \\
   \hline
   15  & 3 & 36 & 5 & 26  \\
  \hline     
   \end{tabular}
   \caption{\label{tab:caracgrid1} Grid characteristics~: $M$~: mesh
     size, $e$~: grid bar thickness, $G$~: solidity, $f$~: frequency, $S$~:
     stroke.} 
 \end{table}
 
\subsubsection{Expected flow characteristics}
From the chosen grid parameters, it is possible to estimate the
expected flow characteristics, by means of empirical laws.
The integral length scale $L$ increases linearly with the distance to the grid $z$:
\begin{equation}
  L~=~c_Lz
  \label{eq:intL}
\end{equation}
where $c_L$ is a constant that depends upon the grid geometry. For comparison we will use $c_L=0.2\pm 0.05$, which is the values \textcite{Hopfinger1976} obtained with $S/M = 8/5$, the closest to our configuration.

For simplicity, we define the origin of the vertical coordinate
$z$ as the midpoint of the oscillation even though \textcite{Hopfinger1976} report virtual origins of order $M$.

The transverse (horizontal) fluctuating velocity $\sigma_u$ has been shown to follow: 
\begin{equation}
  \sigma_u~=~c_ufM^{1/2}S^{3/2}z^{-1}
  \label{eq:sigmaU_OG}
\end{equation}
where $c_u$ is a constant that depends upon the grid
geometry. Based on the literature~\cite{Hopfinger1976, DeSilva1992}
we consider $c_u=0.25$. The fluctuating velocity decreases as the
inverse of the distance $z$.  

In the sequel we will also measure the dissipation rate per unit mass of the
flow $\epsilon$. In previous grid experiments, it has been shown to
behave as:
\begin{equation}
\label{eq:epsilon_L}
  \epsilon~=~C_\epsilon~\frac{\sigma_u^3}{L}
\end{equation}
where $C_\epsilon~\approx~1$ \cite{Sreenivasan84}.

Assuming that the flow is quasi homogeneous and isotropic, from the
dissipation rate one can then infer the Kolmogorov dissipative length scale 
\begin{equation}
  \eta~=~\left(\frac{\nu^3}{\epsilon}\right)^{1/4}
\end{equation}

It is generally unclear how exactly the kinematic viscosity $\nu$ should be defined in
He~II flows. \textcite{Babuin14} have measured an effective viscosity $\nu_\text{eff}$,
defined as the ratio between the dissipation rate and the mean enstrophy,
in a turbulent grid flow. They found that around 2~K, the value of $\nu_\text{eff}$
is of the same order of magnitude as $\mu_n/\rho$ where $\mu_n$ is the dynamic
viscosity of the normal component. In the sequel, we thus use
$\nu=\mu_n/\rho\approx 1\times 10^{-8}\,\mathrm{m^2/s}$ as an approximation in He~II.

Table~\ref{tab:caracgrid2} summarizes the above primary and derived quantities in our experimental conditions.

\begin{table}[ht!]
  \begin{center}
    a)
    \begin{tabular}{|c|c|} 
      \hline
      $L$ & $\sigma_u$\\
      \hline
      [mm] & [mm/s]\\
      \hline
      $14\pm5$& $9.3_{-1.9}^{+2.6}$\\
      \hline     
    \end{tabular}\\
    b)
    \begin{tabular}{|c|c|c|c|c|c|} 
      \hline
      T   & P     & $\epsilon$            & $Re_\lambda$ & $\eta$   &$\tau_\eta$\\
      \hline
      [K] & [bar] & [$10^{-5}\mathrm{m^2/s^{-3}}$] & [-] & $\mu$m   & [ms]      \\
      \hline
      2.8 & 1     & $5.8_{-3.7}^{+12.2}$    & 280        &  22      & 20        \\
      \hline
      3.5 & 1     & $5.8_{-3.7}^{+12.2}$    & 270        &  24      & 21        \\
      \hline
      2   & 0.031 (sat)  & $5.8_{-3.7}^{+12.2}$    & 440        &      -   &  -        \\
      \hline     
    \end{tabular}   
  \end{center}
  \caption{Expected flow characteristics.
  a) Primary quantities obtained from correlations[\ref{eq:intL},\ref{eq:sigmaU_OG}]  at $f=5Hz$,
  $\frac{S}{M}=1.77$, $\frac{z}{M}=4.6$. The main assumption is that the energy injection at large scale does not depend on the fluid. The
  uncertainties are computed considering min and max values of $z$
  in the field of view together with reported uncertainties on $c_L$ and 
  $c_u$.\\
    b) Derived quantities: $\epsilon$ from eq.~\ref{eq:epsilon_L}, $Re_\lambda$ : Reynolds number based on the Taylor length and is obtained under the assumption of homogeneity and isotropy of the flow as $Re_\lambda= \sqrt{15L\sigma_u/\nu}$, $\eta$
    Kolmogorov length scale, $\tau_\eta$~: Kolmogorov time scale.} 
  \label{tab:caracgrid2}
\end{table}

\subsection{Cryostats}
Usually, visualization experiments in cryogenic facilities are
performed in stainless steel cryostats with small planar optical
accesses to minimize heat losses in the Helium bath \cite{Bewley08,
  LaMantia12, Rousset01}. However, the field of view is then limited
by the number of available windows and by the diameter of the
windows. We have chosen for our facility to use glass (rather than
stainless steel) as material for the cryostat, in order to have a
higher level of versatility for visualization purposes. For instance,
although we only present here 2D measurements, a glass cryostat with
full optical access, allows to consider in future campaigns
multi-camera experiments for simultaneous recordings at several viewing angles,
hence allowing well resolved 3D particle tracking
measurements. The use of a glass cryostat has the additional benefit
of being less expensive than the mixed stainless steel/glass solution,
as it avoids the requirement of sophisticated welding between a
stainless-steel cryostats and optical accesses. 

Two cylindrical concentric double-wall glass vacuum cryostats are
used. The inner cryostat contains the liquid Helium bath, where the
turbulence is generated. The outer cryostat contains liquid nitrogen,
and plays the role of thermal shield to limit losses between room
temperature and the bulk of liquid Helium. Glass is naturally opaque
to the infrared radiations and is heated by room temperature
radiations and in turn heats up the liquid nitrogen hence producing
bubbles. These bubbles disturb the visualization through the
cryostats. To avoid this perturbation, the level of nitrogen is kept
below the visualization area during operation of the experiment. This
in turn reduces the efficiency of the Nitrogen thermal shield. To
further minimize radiation heat load, an aluminum shield (item 8 in
fig.~\ref{fig:PresOgres}-a) is also
immersed inside the liquid nitrogen cryostat. Holes are made in the
aluminum shield at the level of the visualization area. Note that this
aluminum shield is disposable and a new adapted shield can easily be
prepared if cameras are added to the experiment or if the
visualization area needs to be enlarged.\\

\subsection{Visualization system}
\begin{figure}[ht!]
\setlength{\unitlength}{1cm}
\begin{center}
\includegraphics[]{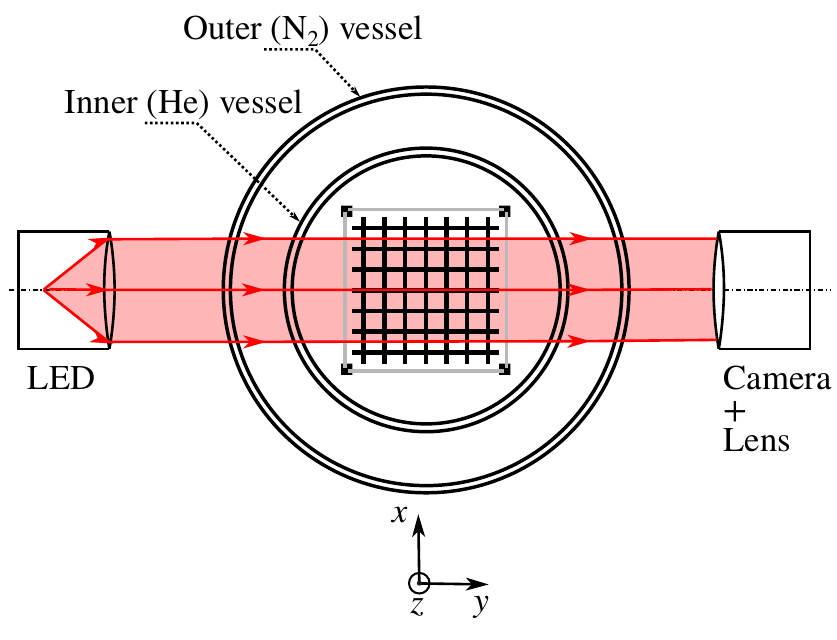} 
\end{center}
\caption{Top view of the visualization system.}
\label{fig:LED}
\end{figure}

Measurements are based on high-speed visualization with a Phantom V12
camera (with a maximum frame rate of 6200 images per second at the
highest resolution of 1280~pixels~$\times$~800~pixels on a one inch CMOS
sensor). We use a red Light Emitting
Diode (LED) with a collimation lens in order to produce an
approximately parallel light beam aiming straight on the camera lens
as shown in fig.~\ref{fig:LED}. 

In this backlight configuration, we record the shadows of the
particles travelling across the light beam, with an \textit{a priori}
undetermined position in the $y$ direction (along the line of sight of
the camera).  We therefore use a lens with a large numerical
aperture (a long distance microscope K2-SC CF-1/B) that ensures both a
good luminosity (and contrast) and a small depth of field. The latter has
been measured to be of the order of $\delta_{dof}\simeq 1.4$~mm, hence ensuring a quasi-2D
measurements at a fixed known $y$-position. Besides, the backlight
configuration allows for a good contrast with a low light power,
which minimizes the heat sources into the helium bath.

All the measurements discussed here have been done at a vertical
position centered around a distance $z = 4.6\,M$ below the average position
of the grid. The overall field of view is $1.8M \times 1.2M$
(\emph{i.e.} $25 \times 18$~mm$^2$). 
The camera lens was located 
at a working distance of \SI{35.5}{cm} from  the center of the aquarium.
The geometrical configuration is the same in He~I and in
He~II.

\subsection{Particles}
Particle seeding was done using K20 type hollow glass microspheres
from 3M. Those particles are commercially available as a poly
disperse population with sizes ranging from 10 to \SI{200}{\mu m} and
densities ranging from 130 to \SI{200}{kg.m^{-3}}.

We first sieved those particles in order to remove the largest and the
smallest ones. Only particles that have a diameter larger that
\SI{71}{\mu m} and smaller than \SI{100}{\mu m} have been retained for our
measurements. The particle size 
distribution has then been measured using a Spraytech diffractometer
from Malvern Instruments Ltd. The particle mean diameter D32 (defined
as the ratio between the mean volume and the mean area) has been found
to be of the order of \SI{85}{\mu m}.

We also measured the average particle density: we immersed a know mass
of particles in  a known volume of water, and measured the resulting
total volume. We found a mean particle density of the order of
\SI{177}{kg.m^{-3}}.

Table \ref{tab:caracpart} summarises the particle characteristics. \\
\begin{table}[ht!]
   \centering
   \begin{tabular}{|c|c|c|c|c|c|c|c|c|c|c|} %|l|l|l|l|l|l|l|l|l|l|l|
   \hline
 Material & $\Phi_p$ & $\rho_p$ \\
   \hline
   [-] & [$\mu$m] & [kg/m$^3$] \\
   \hline
   hollow glass & 85~$\pm$~15  & 177~$\pm$~45\\
   \hline     
   \end{tabular}
   \caption{\label{tab:caracpart} Particles characteristics~: $\Phi_p$~: mean diameter, $\rho_p$~: mean density}
\end{table}

Finally, we observed the shape of the particles with a binocular and verified that they were spherical except for a small fraction corresponding to broken particles. These pieces of sphere were no longer hollow and contributed to a slight increase in overall density, so that the value of \SI{177}{kg.m^{-3}} is a maximum value. During the experiment, the broken particles sank rapidly after injection. 

Particles are dried and injected in the flow using a removable
cryogenic syringe. A few minutes before recording data, we start
oscillating the grid. This has two main advantages: (i) the flow
reaches a steady state and (ii) dense or broken hollow micro-spheres settle and only
particles with a density close to the density of the fluid stay in our
visualisation field. Typically we estimate that the difference between particles
density and fluid density is lower than 15\%.

\section{\label{sec:procedure}Experimental Procedure}
As mentioned in the introduction, we aim at performing experiments
both in normal He~I and superfluid He~II. From a cryogenic point of
view, a fundamental difference between these two states of liquid
Helium concerns the heat conductivity. While He~I has a very low thermal
conductivity (e.g. \SI{0.02}{W.m^{-1}.K^{-1}} at \SI{3}{K} and \SI{1}{bar}),
He~II has a very high effective thermal conductivity. As a
consequence, a He~II bath is 
quasi isothermal, as opposed to a He~I bath.

If the free surface of liquid Helium is at saturation pressure
$P_\text{sat}(T_\text{free-surf})$, in absence of a temperature gradient in the fluid, any point
below the free surface is sub-cooled  due to the pressurization
resulting from the immersion depth. In He~II it is reasonable to assume
there is no temperature gradient and the liquid is therefore always
sub-cooled. This ensures that no bubbles can appear and perturb the
flow.

On the contrary, He~I has a low thermal conductivity and the
temperature of the liquid below the free surface can  increase due to
parasitic heat inputs (through the walls). Those temperature differences
can easily overcome the sub-cooling due to the immersion depth, resulting in
boiling inside the bath.

In order to avoid the presence of bubbles in the field of view, the following
procedure is applied for He~I experiments. After filling with liquid
helium, the bath is cooled down to 2.4K by pumping, the liquid level
being kept well above the top of the test section.  Then, helium
gas at atmospheric pressure is reintroduced above the liquid interface
enabling pressurization and stratification (cold liquid is denser and
remains at the bottom). This pressurization process gives enough time
to perform quasi-stationary measurements in He~I 
without bubbles: we can typically have half an hour at the
operating grid frequency (5~Hz) before the temperature at the
visualization level reaches the saturation temperature
($\approx\SI{4.2}{K}$) generating bubbles again.

In He~II, a MKS 600 valve is used to control the bath pressure
(hence the temperature).  As previously mentioned,
the grid is oscillated a few minutes before taking measurements in
order to ensure a steady state is reached \cite{McKenna2004}. For the
three explored experimental condition (see tab.~\ref{tab:confexp}), at
least 80 films of 400 images are 
recorded to achieve a good statistical convergence. To resolve
particle dynamics, the sequences of 400 images are recorded at a
frame rate $F_s = 3000$~frames per second so that $\delta t
\ll \tau_{\eta}$ where $\tau_{\eta}$ is the dissipative Kolmogorov
timescale of the flow (previously estimated in
table~\ref{tab:caracgrid2}) and $\delta t = F_s^{-1}$ the time between
two images. Typically we have 60 frames per $\tau_\eta$.

Furthermore, images extracted from different films can be considered as uncorrelated
as the delay between two consecutive films is \SI{20}{s}, which is greater than the 
integral time of the flow (\SI{1.4}{s}) 

The different test conditions explored in this paper consist in three
different different configurations that are summarized in
table~\ref{tab:confexp}.

\begin{table}[ht!]
   \centering
   \begin{tabular}{|c|c|c|c|c|c|c|c|c|c|c|} %|l|l|l|l|l|l|l|l|l|l|l|
   \hline
     Config. & Fluid & $T$ & $\rho_{f}$ & $f$ \\
     \hline
     [-] & [-] & [K] & [kg/m$^{3}$] & [Hz] \\%& [-] \\
     \hline
     1 & He~I & 2.8$\pm0.1$ & 145.0 & 5 \\%& \color{red}$\blacksquare$ \\
     \hline
     2 & He~I & 3.5$\pm0.1$ & 138.0 & 5 \\%& \color{orange}$\bullet$ \\
     \hline
     3 & He~II & 2        & 147.5 & 5 \\%& \color{blue}$\blacklozenge$ \\
     \hline
   \end{tabular}
   \caption{\label{tab:confexp} Experimental conditions~: $T$ the
     temperature, $\rho_{f}$ the density of the carrier fluid, $f$ the
     oscillating frequency of the grid.} 
\end{table}

\section{\label{sec:improc}Image processing}
In this section we describe how the image sequences are post processed
to first determine the position of individual particles at a given
time $t$, and then to reconstruct tagged particle tracks along time.

% We have used the information of the detected centers to address two
% important questions of super-fluid turbulence, which are discussed in
% the sequel:  
% \begin{enumerate}
% \item Is the spatial distribution of the particles different in He~II
% compared to He~I?
% \item Is the turbulent dynamics of the particles different in He~II compared to to He~I and in particular, is there any measurable difference in the energy dissipation? 
% \end{enumerate}

% The first item, which is discussed in the next section, only requires
% position statistics (without the necessity to determine particle
% tracks. The second item, requires to access the dynamics of the
% particles and hence to reconstruct the temporal tracks of the
% particles in order to infer particles velocity and eventually their
% acceleration.  The temporal tracking procedure and the subsequent
% results will be discussed in the corresponding sections.

\subsection{\label{sec:detection}Particle detection}

\begin{figure}[ht!]
\setlength{\unitlength}{.1cm}
\begin{center}
\includegraphics[]{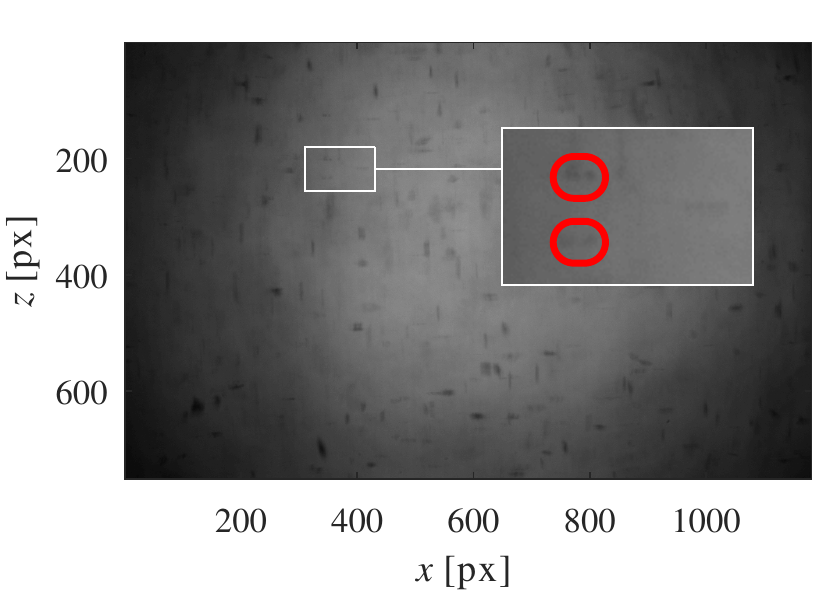}
\end{center}
\caption{Raw Image with a zoom on ghost particles. Field of view of 1280pixel x 800pixel.} 
\label{fig:rawimg}
\end{figure}

Fig.~\ref{fig:rawimg} shows a typical raw image of the hollow
microspheres to be tracked in the oscillating grid flow. This
image shows important optical distortions which may affect particle
detection and eventually the accuracy of the overall particle tracking
procedure. It can indeed be seen that out of focus particles are
strongly distorted, with either a vertical or a horizontal image. This
anisotropic distortion is classical of a cylindrical lens effect, very
likely induced by the cylindrical double walls of the inner and
outer cryostats.  The main goal of the image post-processing is to
correctly detect the particles which are in focus.\\  

\begin{figure}[ht!]
\setlength{\unitlength}{1cm}
\begin{center}
\includegraphics[]{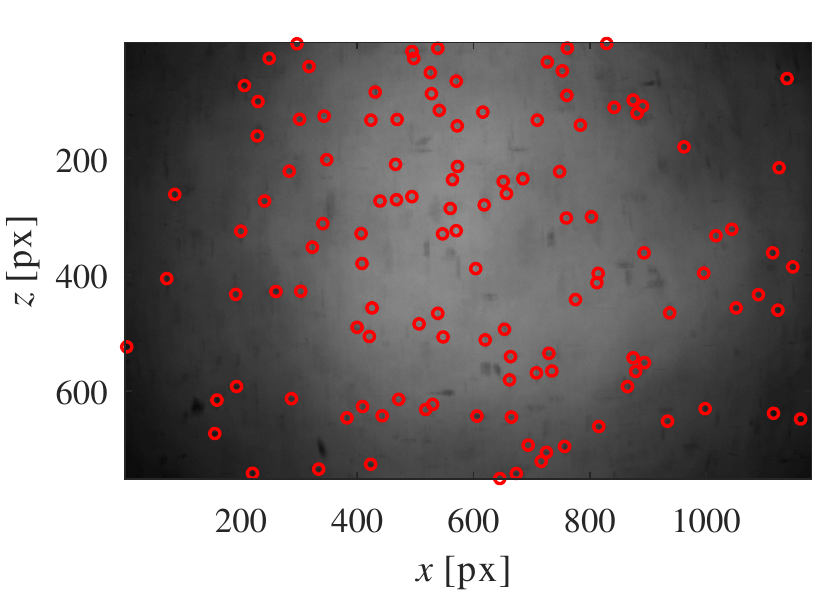} 
\end{center}
\caption{Detection of particles in the raw image from fig.~\ref{fig:rawimg}.}
\label{fig.detection_part}
\end{figure}

The overall image processing sequence is as
follows. First, to clean images,  we apply morphological opening of
the image, in order to retrieve the slightly inhomogeneous background
illumination,  which is subtracted from the corresponding
image. Second, a thresholding is applied to select the most contrasted
particles; this eliminates most particles that are out of the depth
field, as they are dimmer.  Finally, in focus particles are often
found to exhibit a pattern with multiple (typically 3) images closely
aligned in the horizontal direction. This is very likely due to
multiple reflections between the walls of the concentric cryostats. To
remove this effect, a morphological closing using a small horizontal
segment is used to connect dark pixels (we recall that particles
appear dark on a bright background) that are closer from each other by
less than 5 pixels (i.e. $\SI{100}{\mu m}$).  This leads to processed image were most in focus
particles appear as smooth blobs of pixels. Their center is determined
as the center of mass of these blobs, see fig.~\ref{fig.detection_part}.

\subsection{\label{sec:lpt}Particle Tracking}
Once particles are identified (typically we have an average of 80 particles per image),
they are tracked to get their trajectory along time. For this 
we perform Lagrangian Particle Tracking using the particle tracking code by 
Nicholas Ouellette\cite{OuelletteTracking}.

\begin{figure}[ht!]
\setlength{\unitlength}{1cm}
\begin{center}
\includegraphics[width=8.8cm]{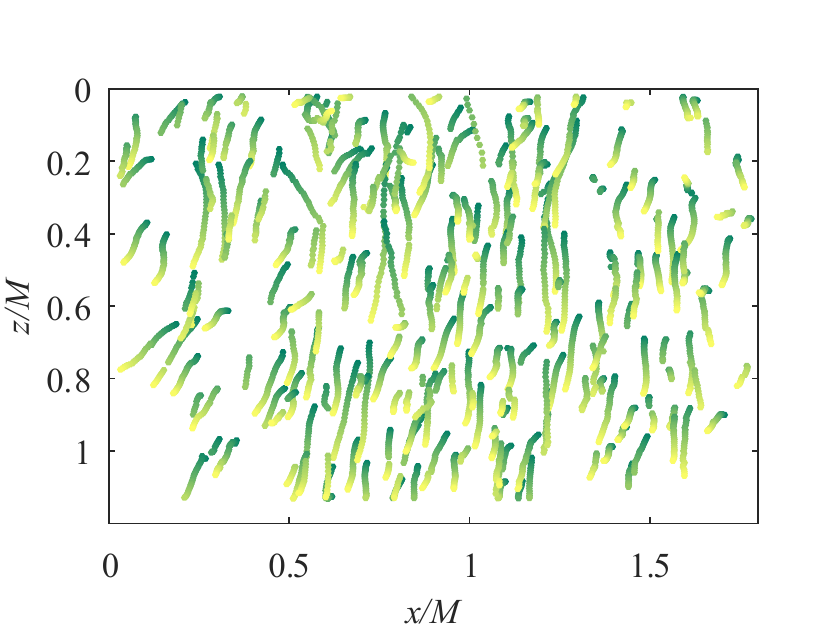} 
\end{center}
\caption{Particle trajectories over 400 images in He~II (config.~3 in tab.~\ref{tab:confexp}). Dark green is for t~=~0 and light green relates to the end of a video (400$^{th}$ image) at t~=~133ms.}
\label{fig:trajpart}
\end{figure}

In fig.~\ref{fig:trajpart}, we show typical trajectories obtained
after particle tracking over a sequence of 400 images.

Lagrangian velocity and acceleration are obtained by convolution of
the raw trajectories with a truncated Gaussian smoothing and
differentiating kernel~\cite{Mordant04} to filter high frequency noise
from the recorded trajectories. Traditionally, the width of the
filtering kernel is chosen as to minimize the impact of noise on
acceleration
variance~\cite{,bib:mordant2004_PRL,Ouellete06b,Ouellette06,bib:klein2013_MST}. However,
the level of small scale noise is particularly high in our experiment
compared to classical experiments at ambient temperature, due to the
multiple curved optical interfaces between the core of the cryostat
and the cameras. As a consequence, the typical time scales of the
noise, overlap with the small turbulent dynamics of the particles and
estimates of acceleration remain sensitive to the choice of the
filtering width, what affects the robustness of acceleration
statistics (see section~\ref{sec:epsiloneta}). Future experiments are planed to
improve this, by an entirely new design of the cryostat in order to
avoid multiple layers of curved interfaces. For the present study, we
therefore chose the filtering properties based on particles velocity
variance, which is less sensitive to small scale noise than
acceleration. We find that a Gaussian  smoothing kernel of width 6~ms
limits reasonably the impact of noise with a weak impact on velocity estimates.

\begin{figure}[ht!]
\setlength{\unitlength}{1cm}
\begin{center}
\includegraphics[width=8.8cm]{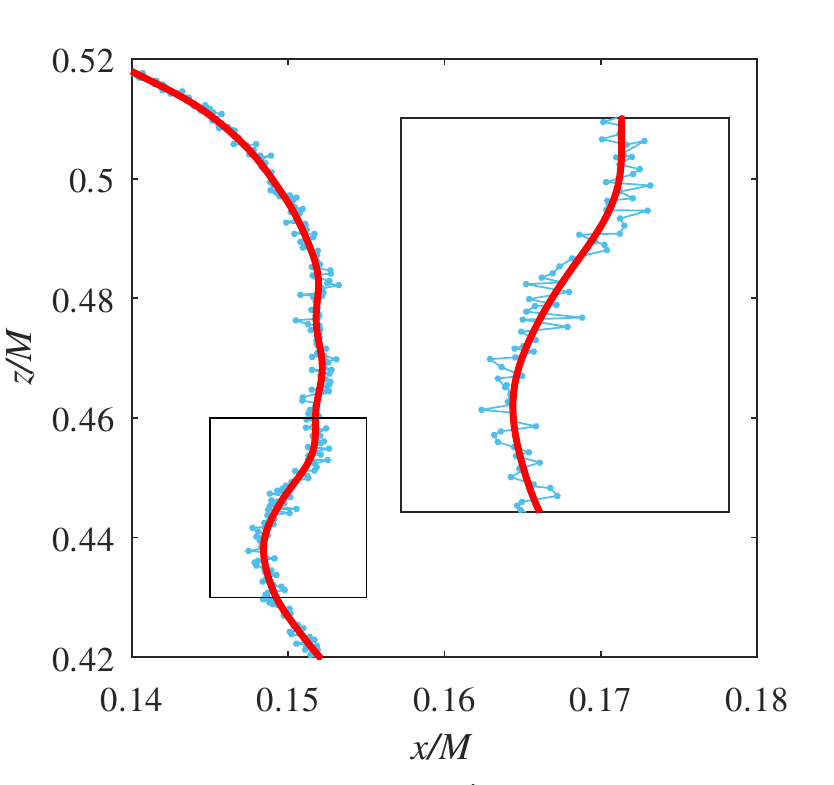} 
\end{center}
\caption{Particle trajectory filtered with a Gaussian kernel of width 18 points (6~ms).}
\label{fig:filteredTrack}
\end{figure}

Fig.~\ref{fig:filteredTrack} represents an example of such raw and
filtered trajectory, clearly showing the high degree of small scale
noise.  
As discussed in sec~\ref{sec:vel}, even after this filtering procedure
velocity increment statistics at inertial scales may still be weakly
affected by some remaining level of noise. This lead us to deploy a
strategy in order to access robust statistical estimate of the
Lagrangian dynamics of the particles based only on their position (and
position increments) statistics, hence avoiding the amplification of
noise associated to numerical  differentiation (see
section~\ref{sec:vel} and section~\ref{sec:methodPPD}). 

\section{Preferential concentration study}
\label{sec:trapping}
The flow conditions explored here lead to a Kolmogorov scale (in He~I) of the order of
$\SI{23}{\mu m}$. According to \textcite{Babuin14}, we expect that the inter-vortex
distance should be of the same order. Thus, the $\SI{80}{\mu m}$ micro-spheres used here
are also of the same order of magnitude. In addition, the microspheres that remain
in the field of view after a certain delay have a density very close to the density
of the fluid. These particles can therefore be expected to behave like tracers and
be randomly distributed in space. In this section, we will verify this in He~I,
a classical fluid. 

Contrary to ideal tracers, which follow the flow and are randomly distributed, 
inertial particles may experience clustering (see
e.g. \textcite{Monchaux12}). On the other hand, even for 
tracer particles, preferential concentration may arise in He~II due to the
trapping of particles about the core of the quantized vortices. This
phenomenon has been widely studied by direct visualisation of
turbulent counter-flow
experiments~\cite{Williams74,Murakami88,Paoletti08,Guo13} in He~II at 
rest, but has never been addressed in mechanically forced  superfluid
turbulence.

\begin{figure}[ht!]
\setlength{\unitlength}{1cm}
\begin{center}
\includegraphics[]{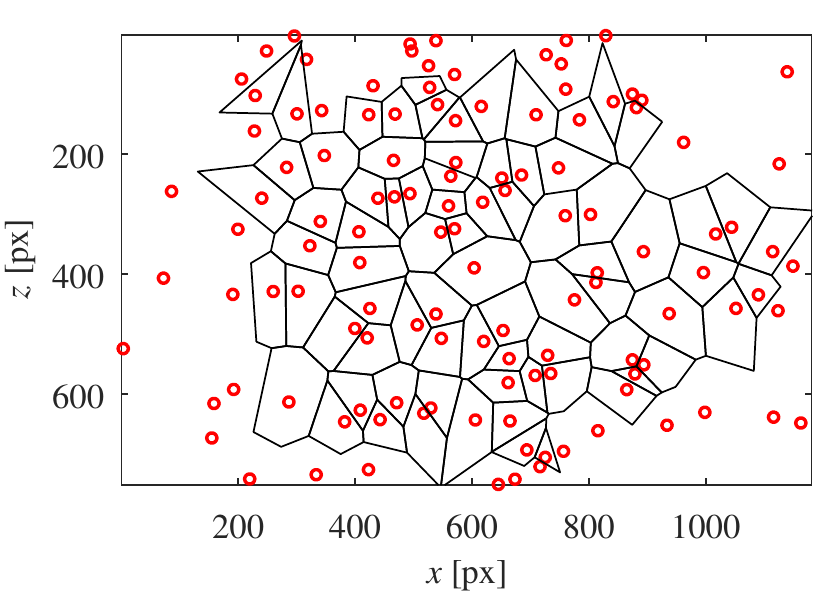} 
\end{center}
\caption{Vorono\"{i} diagram of particles detected in fig.~\ref{fig.detection_part}.}
\label{fig:voronoi_diagram}
\end{figure}

We propose in the sequel an original analysis (based on Vorono\"i
tesselations) of the spatial distribution of the detected particle
centers in order to explore whether particles exhibit some non-trivial
structuration. A Vorono\"{i} diagram consists in defining
a cell  that contains all the points of the space that are closer to
a given particle  than to any other particle. Fig.~\ref{fig:voronoi_diagram} presents the 
Vorono\"{i} diagram of  particles detected in fig.~\ref{fig.detection_part}. 

In regions where there are a lot of particles, Vorono\"{i} cells have a 
small area and where there are few particles, cells are bigger: the inverse of
cell areas reflects the particles concentration. Hence, the cell areas Probability
Density Function provides a quantitative way to describe the degree of
clustering of a set of particles. This method of analysis of particle
preferential concentration has already been widely used but to the
authors' knowledge never on He~II seeding. We applied this method to
our measurements both in He~I and in He~II.

\begin{figure}[ht!]
\setlength{\unitlength}{1cm}
\begin{center}
  \includegraphics[]{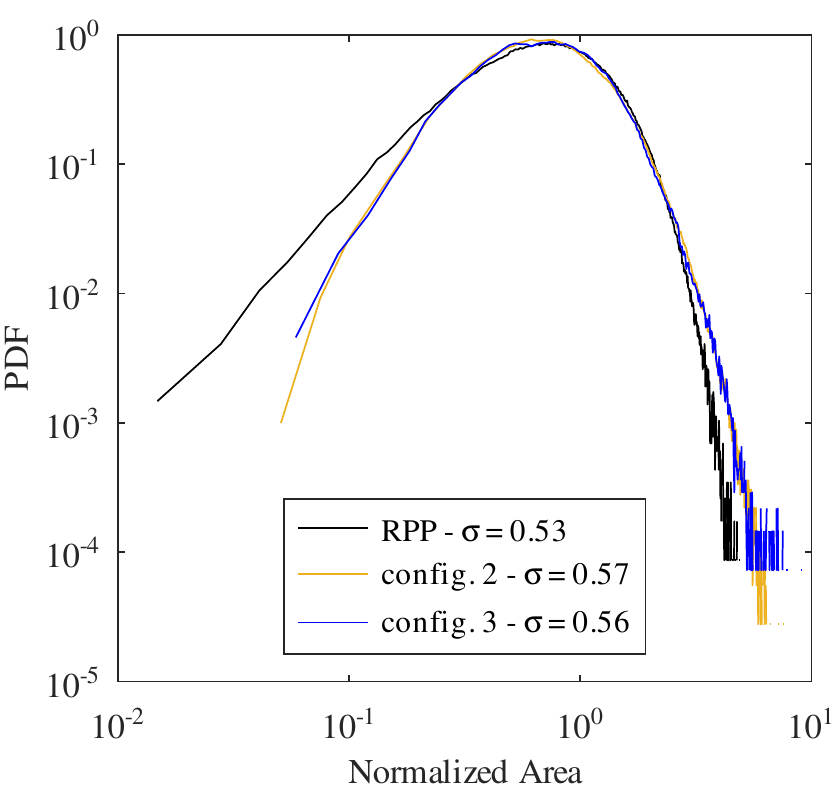} 
\end{center}
\caption{Probability Density Functions of normalized Vorono\"{i} area:
Random Poisson Process (black), config.~2 (orange) and config.~3 (blue).}
\label{fig:Voronoipdf}
\end{figure}

In fig.~\ref{fig:Voronoipdf}, the black curve is the PDF of the
vorono\"{i} cell areas in the case of randomly distributed
particles. This kind of distribution can be modeled by a Random
Poisson Process (RPP) which has a standard deviation
$\sigma\approx 0.53$. The blue and orange curves are the
PDF of vorono\"{i} areas in He~I and He~II respectively. The curves
follow the same trend within the accuracy of our measurements, so the
distribution of particles does not depend on the state of helium.
Furthermore the PDF in He~I  and He~II both have a standard deviation $\sigma$
comparable to that of the RPP. This means that the distribution of the particles
is almost random as expected for a homogeneous seeding and tracer
particles.

We have done further tests that show that the PDF obtained in He~I do not
depend on the height $z$  either.
This demonstrates that the position of the injector is
sufficiently far away to avoid any residual preferential concentration.

% This Vorono\"{i} process provides a robust way of characterizing
% particles by accounting for both the horizontal and the vertical
% components whereas in \cite{Svancara17} the nature of the particles is
% determined from the deviation of the probability density function of
% the Lagrangian velocity increments of one component of the flow.
% Results in He~II follow the same trend as the ones in He~I. 

In conclusion, this Vorono\"{i} diagram method has the double benefit of
analyzing particle trapping and investigating the nature of particles
in the flow.

At this point, it is interesting to have an estimate of the inter vortex line
distance $\delta$. Following \textcite{Babuin14}, we can use evaluate the effective viscosity which leads to $\delta\approx\SI{35}{\mu m}$. 

In our working conditions, i.e. with particles of diameter
$d_p\approx \SI{85}{\mu m}$, slightly larger than the inter-vortex distance, the analysis
provided no evidence for particle trapping in He~II. The particles are
unlikely to be influenced by a single vortex and it would thus
be interesting to study  the case where $d_p \ll \delta$.

\section{\label{sec:velfield} Single Time Velocity Statistics}
Fig.~\ref{fig:trajpart} shows a typical set of reconstructed
trajectories. An overall vertical trend of particles motion can be
seen which suggest the existence of a mean drift velocity,
see~\cite{Eidelman01, McKenna2004}. This is expected as particles are
slightly denser than the carrier fluid, see Table~\ref{tab:carac}.
Besides, to get a zero mean velocity in an oscillating grid
experiment, an infinite aspect ratio of the aquarium for the test
section $H/W$ is required.
% this avoid the impact of any possible recirculation at integral scales. 
In the horizontal direction, where the gravity should have no effect, 
the mean velocity $\langle~u~\rangle$ has been measured
and found to be negligible compared to velocity fluctuations. The
ratio between the average and the standard deviation of the velocity
is $\langle u \rangle/\sigma_{u}\approx 0.1$ in He~I and
$\langle u \rangle/\sigma_{u}\approx 0.3$ in He~II. In the next sections we focus the analysis on the
horizontal component $u$ of the velocity. 

\subsection{\label{sec:pdfvel}Velocity Probability Density Function and classical fluctuating velocity computation} 
\begin{figure}[ht!]
\setlength{\unitlength}{1cm}
\begin{center}
\includegraphics[]{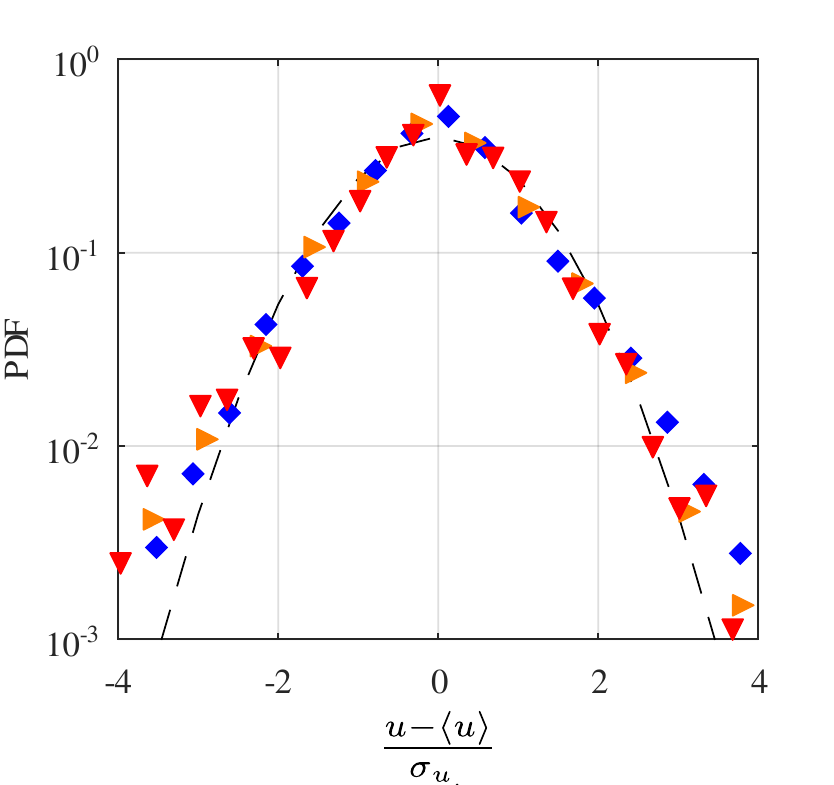} 
\end{center}
\caption{Probability Density Function of normalized velocities alongside the horizontal axis $x$: \color{red}$\blacktriangledown$~\color{black}: config.~1 / $\color{orange}\blacktriangleright$~\color{black}: config.~2 / \color{blue}$\blacklozenge$~\color{black}: config.~3.}
\label{fig:pdfvel}
\end{figure}

Fig.~\ref{fig:pdfvel} shows the probability density function (PDF)
of the horizontal velocity fluctuations. They are found to be
quasi-gaussian, with slightly over-gaussian
tails. Table~\ref{tab:carac} summarises the trends of the standard
deviation of velocity and its comparison with the expected value from
classical empirical laws usually used for oscillating grids in
classical fluids (Eq.~\ref{eq:sigmaU_OG}).  A good agreement is found
with the empirical law, with no major difference between the fluid and
the superfluid cases. A small difference is observed however between
the 2 measurements performed in He~I: config.~1 shows a
larger deviation to empirical laws.

\begin{table}[ht!]
   
   \centering
   \begin{tabular}{|c|c|c|c|c|c|c|c|c|c|c|c|} %|l|l|l|l|l|l|l|l|l|l|l|
   \hline
 Config. & Fluid & $\sigma_u$& $\langle u\rangle$ & $\frac{\sigma_u}{\sigma_{u}^{th}}$ \\
   \hline
   [-] & [-] & [mm/s] &[mm/s] & [-] \\
   \hline
   1 & He~I & 8.3~$\pm$~1.7 & 0.9& 0.86 \\
   \hline     
   2 & He~I & 9.3~$\pm$~1.9 & 0.3& 0.96 \\
   \hline
    3 & He~II & 9.1~$\pm$~1.8 & -3.2& 0.94 \\
   \hline     

   \end{tabular}
   \caption{\label{tab:carac} Summary of horizontal velocity
     measurements obtained using position differentiation.}
\end{table}

\subsection{An alternative way to access the fluctuating velocity }
\label{sec:vel}
Lagrangian velocity is usually obtained from the derivative of
individual particle trajectories (see previous
section~\ref{sec:pdfvel}).  Statistics are then estimated from this
data set of individual velocities.  Such a numerical differentiation
process on individual trajectories tends to amplify the noise
present in the position data of the particles and requires to filter
the trajectories. As described in section~\ref{sec:pdfvel} this is
done here using gaussian-filtering. Choosing appropriate filtering
parameters is not trivial : if individual trajectories are not
sufficiently filtered the statistical quantities estimated (as the
velocity standard deviation) may be biased as they still include noise
contributions, whereas if the trajectories are too filtered, they will
be artificially smoothed.

We propose here an alternative estimation of the standard deviation
of the particle velocity based on purely kinematic considerations of
the position temporal increments $\delta x$. This approach does not
require filtering individual trajectories and hence gives a more
robust estimate. 

Let's consider the mean square displacement of the particles $\delta
x^2=\langle\left(x(t) - x(t+\tau)\right)^2\rangle$ where $x(t)$ and
$x(t+\tau)$ represent the horizontal position at two different times of the same
particles along its trajectory. Assuming the trajectories are smooth
(and differentiable) for sufficiently small timelags while they become
uncorrelated and non-smooth for large timelags, the mean square
displacement is expected to have at least two asymptotic regimes~:

\begin{equation}
  \langle\delta x^2\rangle =  \left\{
    \begin{array}{r c l}
      u_{rms}^2\tau^2 &\,\text{for}\,& \tau~\ll~T_L\\%\text{ r\'egime
      % ballistique (Batchelor
      % \cite{Batchelor50})}
      2u_{rms}^2T_L\tau &\,\text{for}\,& \tau~\gg~T_L %(Taylor \cite{Taylor20})
    \end{array}
  \right.
  \label{eq.DispPair} 
\end{equation}
where $u_{rms}^2$ is the second-order moment of the velocity
and $T_L$ represents the Lagrangian correlation time scale of the
particles motion.  In our study the duration of a video (133ms) is
shorter than the integral time $T_L = 1.4s$, hence only the short term
ballistic regime $\langle\delta x^2\rangle~=~u_{rms}^2\tau^2 $ is
expected to be observed.

In practice, experimental data of particles position do not exhibit a
smooth ballistic regime at the smallest time scales, because of the
presence of experimental noise. This results in a deviation from the
quadratic dependence $\delta x^2\propto \tau^2$ for the smallest
$\tau$. For a purely uncorrelated noise, mimicking perfect brownian
motion at short time scales, one would expect to see $\delta
x^2\propto \tau$. To model the influence of noise, the measured
particle position $x$ can be written as the sum of the real position
$x^*$ (without noise  contribution) and the experimental noise
$\theta$:  $x = x^* + \theta$. The  measured mean square displacement
can then be rewritten, for the short time lags, as 

\begin{equation} 
  \langle \delta x^{2} \rangle = u_{rms}^2 \tau^{2} + 2\langle \theta^2\rangle
  \left( 1 - R_{\theta\theta}\left(\tau\right) \right) +
  \mathcal{O}(\tau^{3}),
 \label{eq.statspos}
\end{equation}
where $R_{\theta\theta}$ is the autocorrelation function of the noise:
\begin{eqnarray}
  \lim_{\tau \to 0}R_{\theta\theta}(\tau) &=& 1,\\
  R_{\theta\theta}(\tau\gg\tau_{\theta}) &=& 0.
  \label{eq:noise}
\end{eqnarray}
Here $\tau_\theta$ is the correlation time scale of the experimental noise.

From eq.~\ref{eq.statspos}, by replacing $u_{rms}^2$ by
$\langle u \rangle^2 + \sigma_u^2$, hence not necessarily assuming the
mean velocity $\langle u \rangle$ is zero, one sees that the standard
deviation of the velocity $\sigma_u$ can be estimated from the
measured mean square displacement: 
\begin{equation} 
\langle \delta x^{2} \rangle - \langle {u} \rangle^2 \tau^2
= \sigma_u^2 \tau^{2} + 2\langle \theta^2\rangle \left( 1 -
  R_{\theta\theta}\left(\tau\right) \right) + \mathcal{O}(\tau^{3}). 
\end{equation}

If we consider time lags $\tau$ sufficiently short to neglect high
order corrections to the ballistic term $\sigma_u^2 \tau^{2}$ (what
implies $\tau \ll T_L$), though longer than the correlation time scale
$\tau_\theta$ of the noise in order to neglect $R_{\theta\theta}$, the
velocity standard deviation can be robustly estimated from simple
finite time position increments (hence without effectively
differentiating the trajectories) from the following relation : 
\begin{equation}
  \sigma_u^2 + 2\frac{\langle \theta^2\rangle}{\tau^2} = \frac{\langle
  \delta x^{2} \rangle }{\tau^2}- \langle {u} \rangle^2,
  \label{eq:estim-sigma-u}
\end{equation} 
for $\tau_\theta \leq \tau \leq T_L$.
The second term on the left hand side can be neglected at large
$\tau$ since the measured total displacement $\delta x$ becomes
much larger than the observed positional noise $\sqrt{\langle \theta^2\rangle}$.

\begin{figure}[ht!]
\setlength{\unitlength}{1cm}
\begin{center}
\includegraphics[]{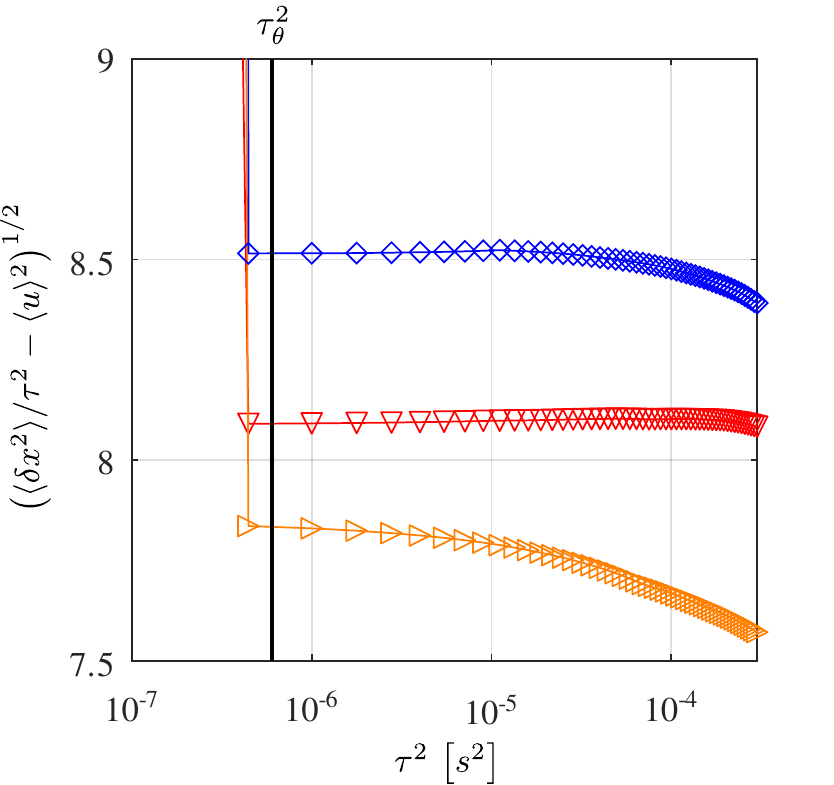} 
\end{center}
\caption{Estimate of $\sigma_u$ from the second-order moment of the separation along
   the horizontal axis $x$ (see eq.~\ref{eq:estim-sigma-u}). \color{red}$\blacktriangledown$~\color{black}: config.~1 / $\color{orange}\blacktriangleright$~\color{black}: config.~2 / \color{blue}$\blacklozenge$~\color{black}: config.~3. f~=~5Hz, $\frac{S}{M}~=~1.77$ and $\frac{z}{M}~=4.6$. The difference in time between each point corresponds to the inter-frame time of the recorded movies.}
\label{fig:mom2posx}
\end{figure}

Fig.~\ref{fig:mom2posx} presents
$\left(\langle \delta x^{2} \rangle/\tau^2 - \langle {u} \rangle^2
    \right)^{1/2}$ as a function of $\tau^2$.  The
rapid initial decrease corresponds to the noise contribution
$2\theta^2/\tau^2$ and possibly also to some reminiscence of the noise
correlation $R_{\theta\theta}$  which may not be exactly zero for the
shortest time lags.
We see however that the curve rapidly reaches a plateau, what suggests
that the contribution of the noise to the position increment variance
vanishes for a time lag corresponding to one or two inter-frame times.

At large time scales, the decrease of the curve correspond to the
onset of high order corrections to the initial ballistic displacement.
The value of the plateau then gives a robust estimate of the standard
deviation of the velocity $\sigma_u$.  These new estimates are
reported in table~\ref{tab.concvit2} for both HeI and HeII experiments
and compared to the estimates from the empirical laws for oscillating
grid turbulence.

It can be noted that this new estimate shows no significant difference
between He~I and He~II. The agreement with the empirical laws is good,
although the measured value is systematically of the order or 20\%
smaller than the empirical estimates. This difference can be
attributed to a slightly different value of the constant $c_u$
(eq.~\ref{eq:sigmaU_OG}) in our experiment compared to tabulated
values in the literature. This may be the consequence of minor
geometrical differences between our setup to the reference ones.

Note that the new estimates of $\sigma_u$ are slightly lower than the direct estimate from Lagrangian velocity. This points to the fact that in spite of the gaussian filtering, taking the derative of the position to estimate velocity remains a noise-amplifying operation. The excess of standard deviation measured from the velocity estimate is very likely due to a choice of filter width too narrow to efficiently reduce the noise.

\begin{table}[ht!!]
   
   \centering
   \begin{tabular}{|c|c|c|c|c|c|c|c|c|c|c|} %|l|l|l|l|l|l|l|l|l|l|l|
   \hline
   Config. & T & $\sigma_{u}$ & $\frac{\sigma_u}{\sigma_u^{th}}$ \\
   \hline
   [-] & [K] & \multicolumn{1}{c|}{[mm/s]} & [-] \\
   \hline
  1 & 2.8 & 8.1 & 0.84 \\
   \hline
  2 & 3.5 & 7.8 & 0.81 \\
   \hline
  3 & 2 & 8.5  & 0.88 \\
   \hline
   \end{tabular}
   \caption{\label{tab.concvit2} Summary of horizontal velocity
     measurements obtained using quadratic displacement fitting.}
\end{table}

\section{Energy budget}
\label{sec:epsilon}
In this section we assess the estimate of the energy injection
rate $\epsilon_{L}$, the energy transfer across inertial scales
$\epsilon_I$ and dissipation rate $\epsilon_{\eta}$.
The energy injection is estimated based on large scale
statistics, using the results of the previous section on velocity fluctuations. 
The energy transfer rate is estimated at inertial scales of turbulence, using the second
order Eulerian structure function and classical Kolmogorov scalings. 
Finally we show an attempt at determining the dissipation rate $\epsilon_\eta$
based on Lagrangian acceleration measurements and use of the
Monin-Yaglom relation, which relates the dissipation rate $\epsilon_\eta$ to
the variance of acceleration. 

In stationary conditions, in classical turbulence, the three estimates are
expected to be identical, as the only channel to dissipate energy is viscosity.
All the injected energy therefore flows across scales via a unique cascade ending
in viscous dissipation. 

In He~II, the question remains somehow open, since other dissipation mechanisms may exist which
would lead to multiple channels for the energy to flow across scales in the normal 
and superfluid components which are eventually coupled via mutual
friction~\cite{Vinen57}. 

It remains unclear at the moment which component the Lagragian particles actually trace in HeII. One goal of the present study is to proceed to different estimates of energy accross scales in order to explore possible deviations to classical behaviors, which may indicate any specificity of superfluid behavior (due either to a preferential sampling of the tracer to one component or the other, or to the existence of different channels for energy to flow and dissipate accross scales). To this end, we have estimated the energy rates at different scales, always assuming fundamental laws as they are known for classical fluid turbulence, seeking scale by scale for significant differences between measurements carried in HeI and HeII.

\subsection{\label{sec:epsilonL}Energy injection at large scales} 
Mechanical energy is injected into the flow at a scale $L$, known as the integral scale of the flow. A fundamental property of classical turbulence, related to the so called \emph{dissipative anomaly} property, relates the energy injection rate $\epsilon_L$ to the standard deviation $\sigma_u$ of velocity fluctuations and to the integral scale $L$ of the flow : $\epsilon_L~=~C_\epsilon~{\sigma_u}^3/L$.

$C_\epsilon$ being a universal constant of order 1~\cite{Sreenivasan98}. In classical fluids, the dissipative anomaly stands for the fact that this relation does not involve viscosity, whereas all the energy which is injected at large scales is eventually dissipated at small scales by viscosity. This implies that dissipation remains finite even in the limit of vanishing viscosity, what in turns implies the appearance of ever smaller scales eventually leading to the energy cascade of turbulence. 

We make use here of eq.~\ref{eq:epsilon_L} to estimate the energy
injection rate. We assume the Reynolds number of our flow is large
enough for $C_\epsilon$ to be constant and take $C_\epsilon = 1$.

We cannot directly estimate at the moment the integral scale of our flow. This would imply measuring Eulerian statistics over a much larger measurement volume than what is currently accessible. We therefore estimate the integral scale based on the empirical law~\ref{eq:intL} for oscillating grids. This is justified due the relative good agreement already reported in the previous section for the fluctuating velocity compared the the corresponding empirical law. Besides, it can be noted that in eq.~\ref{eq:epsilon_L}, the dependency on $L$ is only to the power $1$, while the dependency on $\sigma_u$ is cubic. We therefore expect that major impacts on the overall estimate of $\epsilon_L$ will be associated to changes of $\sigma_u$ rather that eventual small deviations of $L$.  

\begin{figure}[ht!]
\setlength{\unitlength}{1cm}
\begin{center}
\includegraphics[]{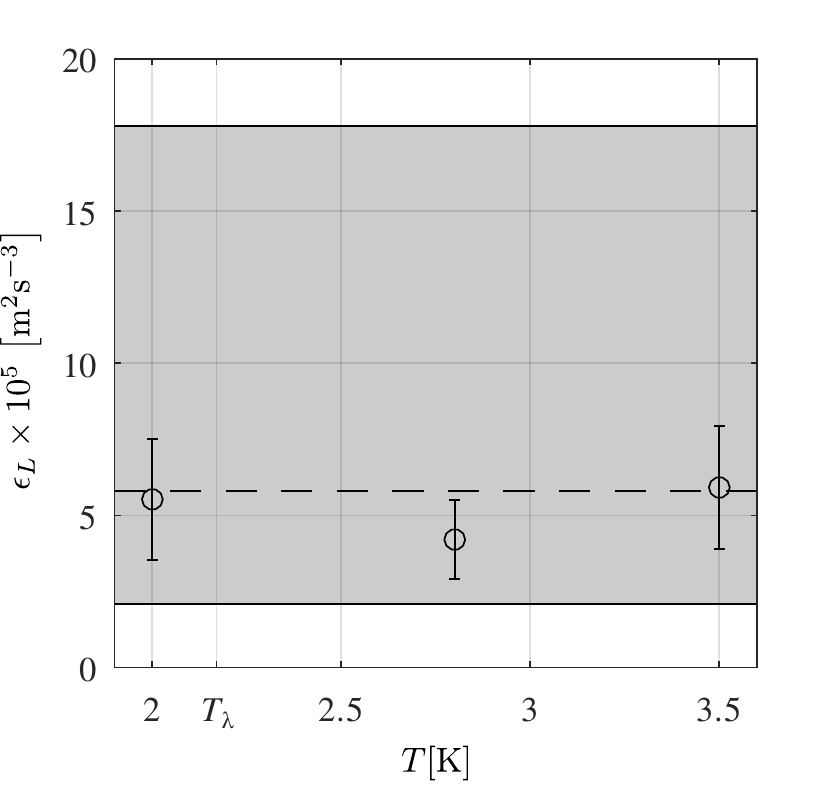} 
\end{center}
\caption{Dissipation rate $\epsilon_{L}$ for the three explored temperatures
computed using the $\sigma_u$ obtained from ballistic separation method.
The gray area shows the range of expected values for $\epsilon$ according to empirical
laws.}
\label{fig:epsilonL}
\end{figure}

Fig.~\ref{fig:epsilonL}, shows the estimates of $\epsilon_L$ for the
three different experimental configurations we have explored. The gray
area in fig.~\ref{fig:epsilonL} depicts the range of expected values
for $\epsilon$ in our experimental conditions, considering the main
uncertainty which lies in the value of experimental constants as
determined by earlier studies $c_u=0.25 \pm 0.025$ and $c_L=0.2\pm
0.05$, \cite{Eidelman01, Fernando94, Villermaux95, Thompson1975,
  Hopfinger1976}.
Furthermore, the experimental area is located at $z=4.6\pm 0.6M$
distance from the grid, and $\epsilon$ is expected to scale as
$z^{-2}$. 

We find that within experimental error bars, the estimates of
$\epsilon_L$ are in good agreement, with empirical laws for classical
turbulence at large scales. Besides, no significant difference is
observed between He~I and He~II.
 
\subsection{Estimate of energy transfer at Inertial scales}
\label{sec:epsilonl}

Assuming classical homogeneous isotropic turbulent scalings, the energy transfer
rate $\epsilon_I$ through the inertial scales is classically estimated
from the Eulerian second-order structure function  
($S_2^E(r) = \frac{11}{3}C_2(\epsilon_I~r)^{\frac{2}{3}}$),
which in a Lagrangian prospect, can robustly be calculated 
from particles relative dispersion statistics~\cite{Bourgoin15}.
As described below, this method has the benefit to give an estimate
of the structure function based on position increments, without requiring the
calculation of particles velocities. This way we avoid having to
differentiate trajectories individually, which as discussed previously
is very sensitive to experimental noise.  

\subsubsection{Methodology - Estimate of Eulerian $S_2^E$ from pair separation statistics}
\label{sec:methodPPD}

The second order Eulerian structure function can be efficiently estimated from
displacement statistics by considering pair statistics.
Particle pair dispersion was first introduced in 1926 by
Richardson~\cite{Richardson1926} and has become since a classical
problem of Lagrangian turbulence. We will only be interested here
in the short time separation regime (also called \emph{ballistic} regime
~\cite{Bourgoin15}), which is the relevant regime to estimate $S_2^E$.
Consider two particles with an initial separation $\vec{D}_0$,
the quadratic relative separation between two particles can be written
$$R_{D_0}^2(t)=\langle|\vec{D}(t)~-~\vec{D}_0|^2\rangle,$$
with $D(t)$ the instantaneous separation between the particles,
and where the average $\langle\cdot\rangle$ is taken over a set of
particles with identical initial separation $\left|D_0\right|$.
By a simple Taylor expansion, one can show that in the limit
$t\rightarrow 0$ (ballistic regime)
$R_{D_0}^2$ is kinematically related to
 $S_2^E$ by:
\begin{equation}
 R_{D_0}^2(t) = S_2^E(D_0)t^2 + \mathcal{O}(t^3).
 \end{equation}

By fitting a quadratic relation for the early stage pair separation
while sweeping the value of initial separation $D_0$, it is therefore
possible to infer $S_2^E(r)$ across scales. Compared to a direct
estimate from the velocity increments, this method to estimate $S_2^E$
has the great benefit to avoid computing position derivative, thus
limiting the amplification of experimental noise. 

Note that in practice, we will only consider the relative quadratic
separation $R_{D_0,x}^2(t)$ in the $x$ direction, so that the above
mentioned procedure will give access to the one-component structure
function $S_{2,x}^E(r)$. In isotropic conditions $S_{2,x}^E(r)$ is simply one
third of the total structure function $S_2^E(r)$).  

Finally, assuming classical Kolmogorov scalings for $S_2^E$, one can
then estimate $\epsilon$ from $R_{D_0,x}^2(t)$ using the relation
(only valid in the limit of small time lags): 

\begin{equation}\label{eq:normR2}
    \frac{R_{D_0,x}^2(t)}{\frac{11}{9}C_2 D_0^{2/3}} =
    \epsilon_I^{2/3}t^2 + {\mathcal{O}}(t^3). 
\end{equation}

\subsubsection{\label{sec:resultsPPD}Results} 

\begin{figure}[ht!]
\centering 
\includegraphics[]{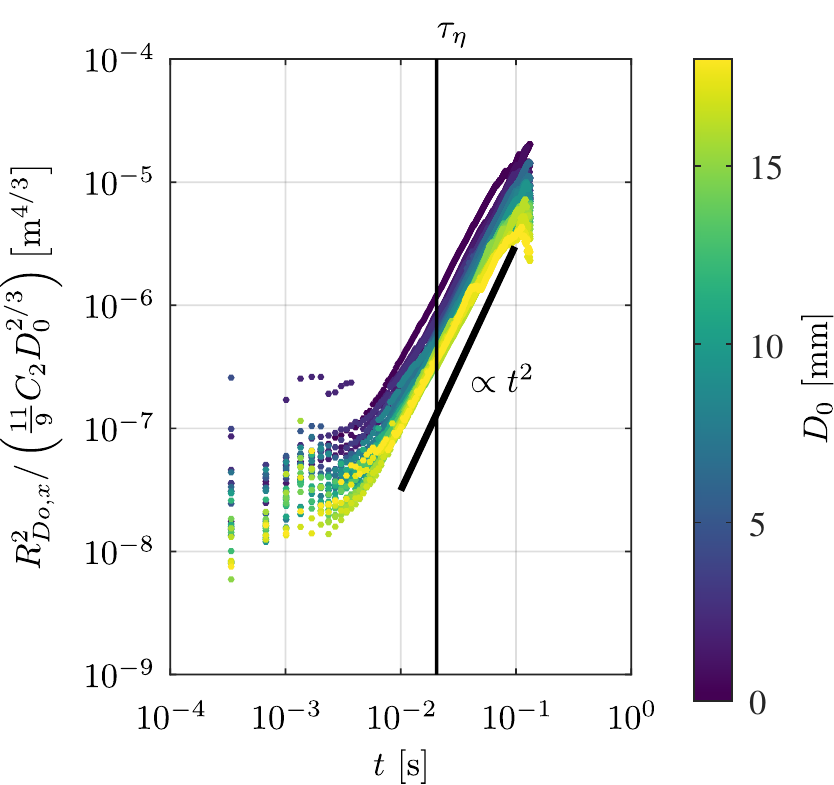} 
\caption{Time evolution of $\frac{R_{D_0,x}^2}{\frac{11}{9}C_2D_0^{2/3}}$
  in config. 3 (He~I).}
\label{fig:seppairesxto}
\end{figure}

Fig.~\ref{fig:seppairesxto} presents the time evolution of the normalized mean
square separation (we only consider the separation in $x$ direction)
$\frac{R_{D_0,x}^2}{\frac{11}{9}C_2D_0^{2/3}}$ against $t$.
Each curve is for a specific bin of initial separation $D_0$.
The expected ballistic regime is clearly visible for time lags
$t\approx\tau_\eta$. 
A deviation from the ballistic regime can be seen at the shortest time lags.
This is a signature of the noise in the position measurement:
in the limit of a purely random (Brownian like noise) the particle
separation rate 
is expected to be purely diffusive ($R^2(t) \propto t$), which is consistent
with the less steep slope at short times.

\begin{figure}[ht!]
\centering 
\includegraphics[]{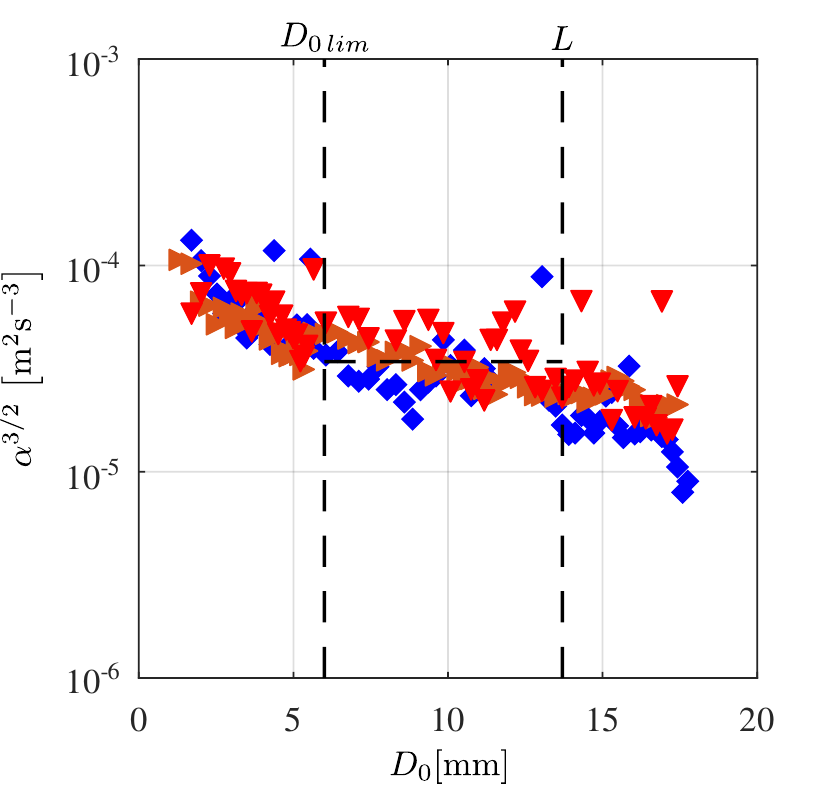} \\
\caption{Coefficient $\alpha^{3/2}$ against the initial separation $D_0$,
    where $\alpha$ is defined as
    $\frac{R_{D_0,x}^2}{\frac{11}{9}C_2D_0^{2/3}} = \alpha t^2+\beta$.
     \color{red}$\blacktriangledown$~\color{black}config.~1,
    \color{orange}$\blacktriangleright$~\color{black}config.~2,
    \color{blue}$\blacklozenge$~\color{black}config.~3. }
\label{fig:epsXppd}
\end{figure}

By individually fitting each curve in fig.~\ref{fig:seppairesxto} against $t^2$,
we can extract the value of the slope $\alpha (D_0) = S^E_{2,x}/\frac{11}{9} C_2D_0^{2/3}$ 
for each initial separation $D_0$.
Assuming Kolmogorov scaling, we can then assess $\epsilon_I$ using the relation
$\epsilon_I = \alpha (D_0)^{3/2}$ (see eq~\ref{eq:normR2}).
The result of this fitting procedure is shown in fig.~\ref{fig:epsXppd}
as a function of the initial separation $D_0$.

A first important interesting finding is that fig.~\ref{fig:epsXppd}
does not highlight any measurable difference between He~I and He~II
situations. We see a pseudo-plateau in the inertial range (we recall
that in the present situation the integral scale is estimated to be
$L\simeq14$~mm), indicating a reasonable Kolmogorov scaling. At small
scales, one would expect a trivial dissipative scaling $S_2(D_0)
\propto D_0^2$, and hence $\alpha \propto S_2/D_0^{2/3}\propto D_0^{4/3}$,
what is clearly inconsistent with the rapid increase observed in
fig.~\ref{fig:epsXppd} as $D_0$ decreases below $D_0\sim 5$ 
~mm. This is primarily due to the fact that small scales are biased by
the finite depth of field ($\delta_{dof}\simeq 1.4$~mm) of our
measurement volume, since we only have access to 2D
measurements. Estimate of separation data are therefore  accurate only
in the limit $D_0 > \delta_{dof}$. Besides, given the dilute nature
of our flow there is not enough statistics at small $D_0$. For
separations $D_0$ of order and smaller than $\delta_{dof}$, the
possible overlap of the 2D projection of particles within the depth of
field, allows for large relative velocities even in the limit of small
apparent separations. 
    
We therefore estimate the dissipation rate $\epsilon_I$, by averaging
$\alpha^{3/2}$ over the
pseudo-plateau in the range 6~mm$< D_0 <$ 14~mm, corresponding to
inertial scales not significantly affected by the finite depth of
field bias. This corresponds to the two vertical dashed lines
represented in fig.~\ref{fig:epsXppd}. 

\begin{figure}[ht!]
\centering 
\includegraphics[]{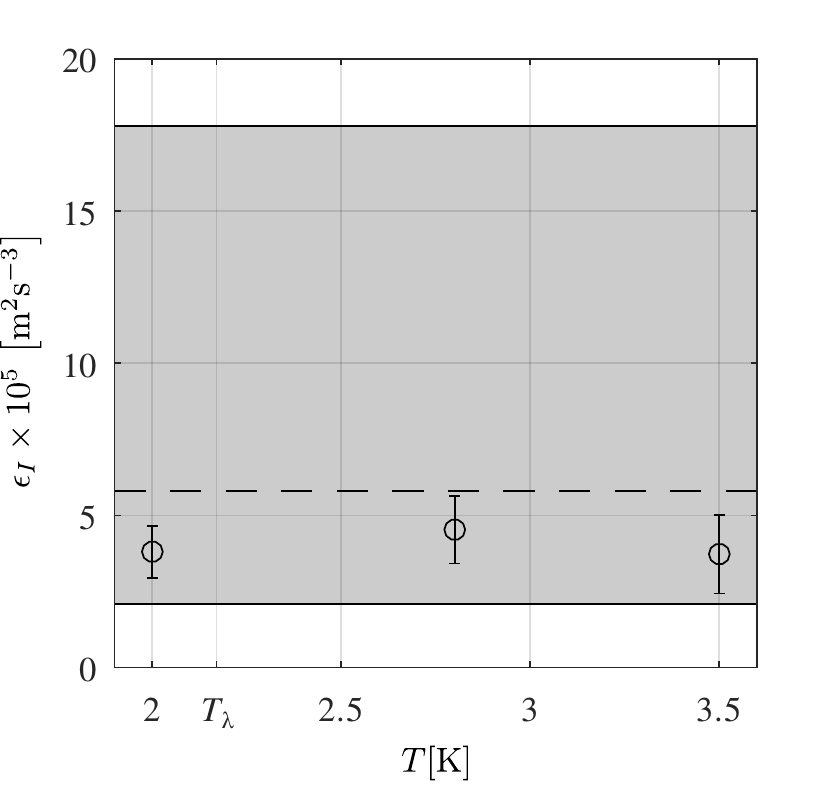} \\
\caption{Dissipation rate $\epsilon_{I}$ for the three explored temperatures
computed the second order structure function estimated from pair separation.
The gray area shows the range of expected values for $\epsilon$ according to empirical
laws.}
\label{fig:epsX}
\end{figure}

The corresponding values of $\epsilon_I$ as a function of the
operating temperature is shown in fig.~\ref{fig:epsX}. It is found
that, within error bars, the inertial scale estimate $\epsilon_I$ is in
good agreement both with empirical laws and with the large scale
estimate $\epsilon_L$ discussed in the previous section. In
particular, no difference is observed between HeII and HeI. 

\subsection{Dissipative scales}
\label{sec:epsiloneta}

Estimation of the dissipation at dissipative scales requires the analysis of
small scale information. In the Eulerian context, this usually drives
back to the 
definition of dissipation $\epsilon_\eta = 2\nu \Omega^2$ (with $\Omega^2$ the
enstrophy), which in homogeneous isotropic turbulence can simply be
rewritten in terms 
of a single component (say $u$) spatial derivative : $\epsilon_\eta =
15 \nu \partial_x 
u$. This requires measurements with high spatial resolution, allowing
to take well 
resolved spatial derivative of the velocity field. In the context of Lagrangian
measurements, as in the present study, the relevant small scale quantity is the
Lagrangian acceleration  (rather than the velocity gradient), which is related
to the dissipation rate via the Heisenberg-Yaglom relation:

\begin{align}
\sigma_{ax}^2~=a_0\epsilon_{\eta}^{3/2}\nu^{-1/2}.
\end{align}

In this relation $\sigma_{ax}$ is the standard deviation of horizontal
acceleration fluctuations and $a_0$ is a dimensionless coefficient that
is empirically known in classical turbulence (from experimental
and numerical studies, see for instance the review article of
\textcite{Toschi09}) and which is known to depend on the Reynolds
number following an empirical law $a_0 \simeq 0.85R_\lambda^{-.25}$,
see~\cite{Voth02}.

Unfortunately, considering the noise issues previously discussed
regarding the direct estimates of velocity statistics, it is unlikely that our
measurements are sufficiently well resolved at small temporal scales to actually
resolve second order Lagrangian derivatives required to estimate
acceleration, as 
taking second order derivatives is extremely sensitive to experimental noise.

\begin{figure}[ht!]
\centering 
\includegraphics[]{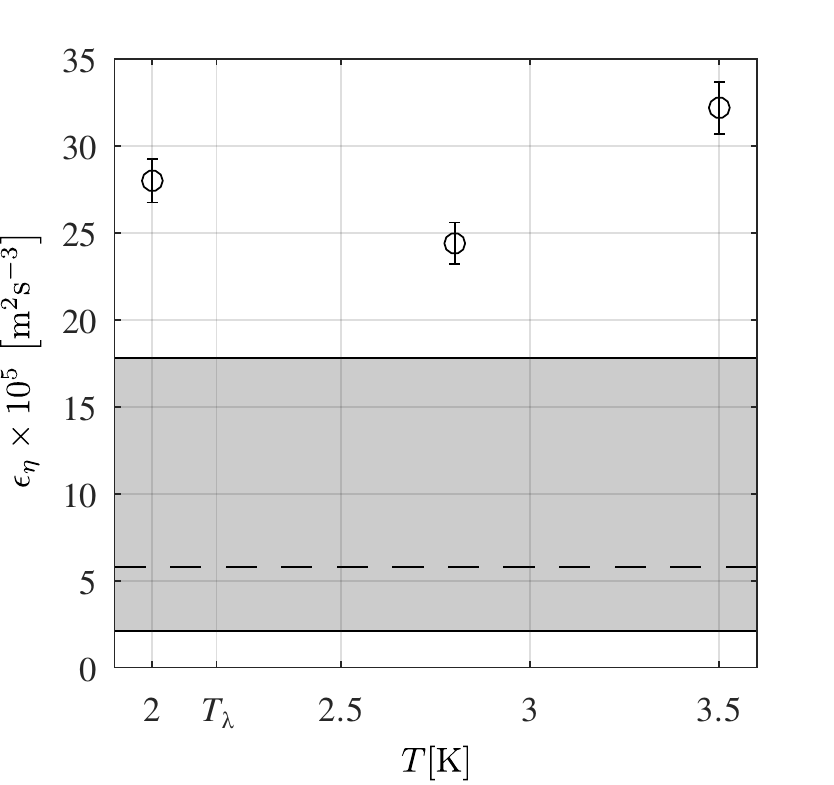} \\
\caption{Dissipation rate $\epsilon_{\eta}$ vs Temperature $T$ at
  dissipative scales.} 
\label{fig:epsHY}
\end{figure}

Still we attempted to perform this estimate. The acceleration is calculated by
convolution of particles trajectories with a second derivative gaussian kernel,
as classically done in Lagrangian studies to estimate filtered
derivatives~\cite{Mordant04}, with the same filtering parameters
(in particular the same filter width as for the direct velocity
estimates previously discussed in section
~\ref{sec:velfield}). 
From this data we calculate the acceleration variance $\sigma_{a_x}$ which
is then used to estimate the small scale dissipation rate from the
Heisenberg-Yaglom relation.
The corresponding results are plotted as a function of the operating
temperature in  fig.\ref{fig:epsHY}.

As one can see on fig.~\ref{fig:epsHY}, our experimental data
do not match the expected values from literature, neither in He~I nor in He~II. 
Consequently, this result cannot be attributed to a particular behavior of HeII
but rather to an insufficient temporal resolution to accurately
estimate the acceleration at dissipative scales.
Efforts are been put at present on the experimental side in order to
improve this and perform new measurements with sufficient resolution at inertial
and dissipative scales.

\section{\label{sec:conclusion}Conclusion}
Our experiment has been calibrated and validated against canonical
oscillating grid turbulence in classical flows. Indeed, our results in
He~I match empirical data.\\ 

We did not find any difference in preferential concentration once HeII
was used. This may be due to the relatively large particle size that
we used (85~$\mu$m) which prevent particles from getting trapped by
superfluid vortex lines. Another possibility is that in isothermal
HeII flow, at sufficiently high Re, the superfluid component is really
linked to the normal one as it has been recently reported that
superfluid component mimics the normal flow, for both large velocity
field and vorticity, see~\cite{Rusaouen17}. \\ 

We noticed the existence of a mean drift of the
particles in the z-direction with a slight impact on the measured
fluctuating velocity in the z-direction. So for the assessment of the
dissipation rate we focused on the horizontal components. At large and
inertial scales results in He~I are as expected in classical
fluids. Furthermore, in these range of currently accessible scales,
our study does not reveal any difference of turbulence properties
between He~I and He~II. This may be explained by the fact that quite
large particles (diameter of 85$\mu$m) are used and at 2K there is
only 39\% of inviscid component in the superfluid. Indeed the effect
of quantum turbulence is expected to appear below a scale where normal
and superfluid component are decoupled (e.g. below the inter vortex
distance). Unfortunately, whilst the spatial and temporal resolution
of our measurements give us access to the dynamics of the flow in the
range of inertial scales, dissipative scales are marginally
resolved. \\ 

Further studies aim at achieving even higher resolution measurements to explore possible differences between classical and superfluid turbulence at and below dissipative scales. To do so, either smaller particles (while keeping spherical shape and monodisperse size) or lower Re number (while remaining fully turbulent) or larger facility should be used.

\bibliography{PRF_OGRES.bib}
\bibliographystyle{plainnat}

\end{document}